\documentclass[12pt,a4paper]{article}

\usepackage{latexsym,amssymb,amsmath,amsbsy,epsfig}

\newcommand{\be}{\begin{equation}}
\newcommand{\ee}{\end{equation}}
\newcommand{\ba}{\begin{eqnarray}}
\newcommand{\ea}{\end{eqnarray}}
\newcommand{\bs}{\begin{subequations}}
\newcommand{\es}{\end{subequations}}
\newcommand{\no}{\nonumber\\}

\newcommand{\diag}{\mbox{diag}}

\begin{document}
\renewcommand{\thefootnote}{\fnsymbol{footnote}}

\title{
\normalsize \hfill CFTP/13-013
\\
\normalsize \hfill UWThPh-2013-14
\\*[7mm]
\LARGE A new $A_4$ model for lepton mixing}

\author{
P.M.\ Ferreira,$^{(1,2)}$\thanks{E-mail: ferreira@cii.fc.ul.pt} \
L.\ Lavoura,$^{(3)}$\thanks{E-mail: balio@cftp.ist.utl.pt} \
and P.O.\ Ludl$ \, ^{(4)}$\thanks{E-mail: patrick.ludl@univie.ac.at}
\\[3mm]
$^{(1)} \! $
\small Instituto Superior de Engenharia de Lisboa \\
\small 1959-007 Lisbon, Portugal
\\[1mm]
$^{(2)} \! $
\small Centre for Theoretical and Computational Physics,
University of Lisbon \\
\small 1649-003 Lisbon, Portugal
\\[1mm]
$^{(3)} \! $
\small Universidade de Lisboa, Instituto Superior T\'ecnico, CFTP \\
\small 1049-001 Lisbon, Portugal
\\[1mm]
$^{(4)} \! $
\small University of Vienna, Faculty of Physics,\\
\small Boltzmanngasse 5, A--1090 Vienna, Austria
\\[3mm]
}

\date{14 October 2013}

\maketitle

\begin{abstract}
We present a new model of the lepton sector
that uses a family symmetry $A_4$ to make predictions for lepton mixing
which are invariant under any permutation of the three flavours.
We show that those predictions broadly agree with the experimental data,
leading to a largish $\sin^2{\theta_{12}} \gtrsim 0.34$,
to $\left| \cos{\delta} \right| \gtrsim 0.7$,
and to $\left| 0.5 - \sin^2{\theta_{23}} \right| \gtrsim 0.08$;
$\cos{\delta}$ and $0.5 - \sin^2{\theta_{23}}$
are predicted to have identical signs.
\end{abstract}

\newpage

\renewcommand{\thefootnote}{\arabic{footnote}}

The experimental discovery that
the lepton mixing angle $\theta_{13}$ is nonzero~\cite{nonzero}
caused a profound change in the subject
of flavour models for the lepton mass matrices.
Many older models ceased to be valid.
New models had to be built;
some recent examples utilizing the horizontal symmetry group $A_4$
are collected in refs.~\cite{flavons,renormalizable}.
Many of those models use `flavons'
and non-renormalizable Lagrangians~\cite{flavons};
in most remaining models~\cite{renormalizable} there are Higgs doublets
at the Fermi scale placed in triplets
of the horizontal symmetry.\footnote{An exception
is the Babu--Ma--Valle model~\cite{mr},
in which the Higgs doublets are $A_4$-invariant.
That model depends on renormalization
to produce realistic neutrino masses and mixings.}
In this paper we present a model that contains only renormalizable terms
and only Higgs doublets which are singlets of $A_4$.

For any $n \times n$ non-singular matrix $M = \left[ M_{\alpha \beta} \right]$,
one may define a matrix $A= \left[ A_{\alpha \beta} \right]$ through
\be
A_{\alpha \beta} = M_{\alpha \beta} \left( M^{-1} \right)_{\beta \alpha},
\ee
where no sum over either $\alpha$ or $\beta$ is implied.
It is obvious from its definition that $A$
satisfies
\be
\sum_{\alpha = 1}^n A_{\alpha \beta} = \sum_{\beta = 1}^n A_{\alpha \beta} = 1.
\ee
The matrix $A$ is invariant under
\be
\label{reph}
M \to X M Y,
\ee
where $X$ and $Y$ are diagonal non-singular matrices.

For our purposes,
$M$ is the (effective) Majorana mass matrix of the three light neutrinos
in the weak basis where the charged-lepton mass matrix is diagonal.
Therefore,
$n = 3$,
the indices $\alpha$ and $\beta$
are in the range $\left\{ e, \mu, \tau \right\}$,
$M$ and $A$ are symmetric,
and $X = Y$ in the transformation~(\ref{reph}).\footnote{The matrix $A$
was also used in this context in ref.~\cite{we}.}
The model in this paper predicts\footnote{Equations~(\ref{pred})
may alternatively be stated as
\be
M_{ee} \left( M_{\mu \tau} \right)^2
= M_{\mu\mu} \left( M_{e \tau} \right)^2
= M_{\tau\tau} \left( M_{e \mu} \right)^2.
\ee
}
\be
\label{pred}
A_{e \mu} = A_{e \tau} = A_{\mu \tau},
\ee
hence
\be
\label{pred2}
A = \left( \begin{array}{ccc}
1 - 2 t & t & t \\ t & 1 - 2 t & t \\ t & t & 1 - 2 t
\end{array} \right),
\ee
where $t$ is in general a complex number.
Moreover,
through the imposition of an additional $CP$ symmetry on our model,
$t$ may be made to be real.
We shall show that
the conditions~(\ref{pred}) fit the experimental data
%Added footnote.
rather well.\footnote{The two conditions~(\ref{pred})
represent a total of \emph{four}\/ constraints
(two from the real parts and two from the imaginary parts)
on the neutrino masses and on lepton mixing.
However,
because those conditions implicitly involve
the Majorana phases of the neutrinos,
which are unobservable in oscillation experiments,
the predictive power of our model is less
than these four constraints might suggest.}

Our model has the usual Standard-Model leptonic multiplets
$\alpha_R: \left( \mathbf{1}, -1 \right)$ and
$D_{\alpha L}: \left( \mathbf{2}, -1/2 \right)$.\footnote{The boldface number
inside each parentheses
is the dimension of the gauge-$SU(2)$ representation;
the second number is the weak hypercharge.}
It has,
besides,
three right-handed neutrinos $\nu_{\alpha R}: \left( \mathbf{1}, 0 \right)$.
The scalar sector is composed of three Higgs doublets
$\phi_k: \left( \mathbf{2}, 1/2 \right)$,
where $k \in \left\{ 1, 2, 3 \right\}$.
Their conjugate doublets are
$\tilde \phi_k \equiv i \tau_2 \phi_k^\ast: \left( \mathbf{2}, -1/2 \right)$.
In our model there are,
besides,
three \emph{real}\/ scalars
$\sigma_\alpha: \left( \mathbf{1}, 0 \right)$.

The model is based on the well-known discrete symmetry group $A_4$
possessing the irreducible representations
\be
\begin{array}{lll}
 \mathbf{1}:\   & S\rightarrow 1,\ & T\rightarrow 1,\\
 \mathbf{1}':\  & S\rightarrow 1,\ & T\rightarrow \omega,\\
 \mathbf{1}'':\ & S\rightarrow 1,\ & T\rightarrow \omega^2,\\
 \mathbf{3}:\   & S\rightarrow \tilde S,\ & T\rightarrow \tilde T,\\
\end{array}
\ee
where
\be
\tilde S =
\begin{pmatrix}
1 & 0 & 0 \\ 0 & -1 & 0 \\ 0 & 0 & -1
\end{pmatrix},
\quad
\tilde T =
\begin{pmatrix}
0 & 1 & 0 \\ 0 & 0 & 1 \\ 1 & 0 & 0
\end{pmatrix},
\ee
and $\omega = \exp{\left( i 2 \pi / 3 \right)}$.
Assigning the fields to the representations as
\be
\left( \begin{array}{c} D_{eL} \\ D_{\mu L} \\ D_{\tau L}
\end{array} \right),
\
\left( \begin{array}{c} e_R \\ \mu_R \\ \tau_R
\end{array} \right),
\
\left( \begin{array}{c} \nu_{eR} \\ \nu_{\mu R} \\ \nu_{\tau R}
\end{array} \right),
\
\left( \begin{array}{c} \sigma_e \\ \sigma_\mu \\ \sigma_\tau
\end{array} \right): \ \mathbf{3}
\ \ \quad \text{and} \quad
\left. \begin{array}{rl}
\phi_1: & \mathbf{1} \\
\phi_2: & \mathbf{1}' \\
\phi_3: & \mathbf{1}''
\end{array} \right.
\ee
leads to the Yukawa Lagrangian
\ba
\label{yukawa}
\mathcal{L}_\mathrm{Yukawa} &=&
- y_1
\left( \bar D_{eL} e_R + \bar D_{\mu L} \mu_R + \bar D_{\tau L} \tau_R \right)
\phi_1
\no & &
- y_2
\left( \bar D_{eL} e_R + \omega \bar D_{\mu L} \mu_R
+ \omega^2 \bar D_{\tau L} \tau_R \right)
\phi_2
\no & &
- y_3
\left( \bar D_{eL} e_R + \omega^2 \bar D_{\mu L} \mu_R
+ \omega \bar D_{\tau L} \tau_R \right)
\phi_3
\no & &
- y_4
\left( \bar D_{eL} \nu_{eR} + \bar D_{\mu L} \nu_{\mu R} 
+ \bar D_{\tau L} \nu_{\tau R} \right)
\tilde \phi_1
\no & &
- y_5
\left( \bar D_{eL} \nu_{eR} + \omega^2 \bar D_{\mu L} \nu_{\mu R} 
+ \omega \bar D_{\tau L} \nu_{\tau R} \right)
\tilde \phi_2
\no & &
- y_6
\left( \bar D_{eL} \nu_{eR} + \omega \bar D_{\mu L} \nu_{\mu R} 
+ \omega^2 \bar D_{\tau L} \nu_{\tau R} \right)
\tilde \phi_3
\no & &
- y_7 \left( \bar \nu_{eR} C \bar \nu_{\mu R}^T \sigma_\tau +
\bar \nu_{\mu R} C \bar \nu_{\tau R}^T \sigma_e
+ \bar \nu_{\tau R} C \bar \nu_{eR}^T \sigma_\mu \right)
+ \mathrm{H.c.},
\ea
where $C$ is the charge-conjugation matrix in Dirac space.
The couplings $y_{1\mathrm{-}7}$ are dimensionless.
There are also bare Majorana mass terms
\be
\mathcal{L}_\mathrm{Majorana} =
- \frac{m}{2} \left( \bar \nu_{eR} C \bar \nu_{eR}^T
+ \bar \nu_{\mu R} C \bar \nu_{\mu R}^T
+ \bar \nu_{\tau R} C \bar \nu_{\tau R}^T \right) + \mathrm{H.c.},
\ee
where $m$ has mass dimension.

When the neutral components of $\phi_{1,2,3}$
get vacuum expectation values (VEVs)
$v_k = \left\langle 0 \left| \phi_k^0 \right| 0 \right\rangle$,
the charged leptons acquire masses given by
\ba
m_e &=& \left| y_1 v_1 + y_2 v_2 + y_3 v_3 \right|, \no
m_\mu &=& \left| y_1 v_1 + \omega y_2 v_2 + \omega^2 y_3 v_3 \right|, \\
m_\tau &=& \left| y_1 v_1 + \omega^2 y_2 v_2 + \omega y_3 v_3 \right|.
\nonumber
\ea
The three quantities $y_k v_k$ must be finetuned
in order that $m_e \ll m_\mu \ll m_\tau$.\footnote{Most models
require a finetuning in order to obtain $m_e \ll m_\mu \ll m_\tau$.
Possible exceptions are models based on the Froggatt--Nielsen
paradigm~\cite{FN} and models with extra dimensions.}

We assume the VEVs of the three $\sigma_\alpha$ to be equal
(see appendix~A):
$\left\langle 0 \left| \sigma_e \right| 0 \right\rangle
= \left\langle 0 \left| \sigma_\mu \right| 0 \right\rangle
= \left\langle 0 \left| \sigma_\tau \right| 0 \right\rangle \equiv s$.
We furthermore assume that $s$ is of the same order of magnitude as $m$,
and that this order of magnitude is very large,
\textit{viz.}\ it is the seesaw scale.
Thus,
the subgroup $\mathbb{Z}_3$ of $A_4$ generated by $T$
is preserved at the high (seesaw) scale
and only gets spontaneously broken at the low
(Fermi\footnote{Another possibility is that
$v_{1,2,3}$ are much lower than the Fermi scale,
if the quarks do not have Yukawa couplings to $\phi_{1,2,3}$
and only couple to an extra doublet $\phi_0$ which is invariant under $A_4$.
In that scheme,
the VEV of the neutral component of $\phi_0$ would be dominant
in giving mass both to the gauge bosons and to the quarks,
while the VEVs of the neutral components of $\phi_{1,2,3}$ would lie much below
the Fermi scale.
The masses of the components of the doublets $\phi_{1,2,3}$
would in this picture lie much \emph{above}\/ the Fermi scale,
due to a type-II seesaw mechanism for Higgs doublets~\cite{radovcic}.})
scale,
when $\phi_2$ and $\phi_3$ acquire VEVs.

The neutrino mass matrices $M_D$ and $M_R$ are defined by
\be
\mathcal{L}_{\nu\, \mathrm{mass}} =
- \bar \nu_R M_D \nu_L - \frac{1}{2}\, \bar \nu_R M_R C \bar \nu_R^T
+ \mathrm{H.c.}
\ee
In our model,
\ba
M_D &=& \diag \left( a,\, b,\,c \right),
\\
M_R &=& \left( \begin{array}{ccc}
m & m' & m' \\ m' & m & m' \\ m' & m' & m
\end{array} \right),
\ea
where
\ba
a &=& y_4^\ast v_1 + y_5^\ast v_2 + y_6^\ast v_3,\no
b &=& y_4^\ast v_1 + \omega y_5^\ast v_2 + \omega^2 y_6^\ast v_3, \\
c &=& y_4^\ast v_1 + \omega^2 y_5^\ast v_2 + \omega y_6^\ast v_3,
\nonumber
\ea
and $m'= y_7 s$.
A seesaw mechanism takes place,
whereupon an effective mass matrix for the light neutrinos
\be
M = - M_D^T M_R^{-1} M_D
\ee
is generated.
The matrix $A$ is then of the form~(\ref{pred2}),
with
\be
t = \frac{{m'}^2}{\left( m' - m \right) \left( 2 m' + m \right)}.
\ee
The values of $a$,
$b$,
and $c$ are irrelevant for $A$.

One may,
if one wants,
furnish our model with an extra $CP$ symmetry,
\be
\label{cpcp}
\begin{array}{rclcl}
& &
\left\{ \begin{array}{rcl}
D_{eL} (x) &\to& \gamma_0 C \bar D_{eL}^T (\bar x),
\\*[1mm]
D_{\mu L} (x) &\to& \gamma_0 C \bar D_{\tau L}^T (\bar x),
\\*[1mm]
D_{\tau L} (x) &\to& \gamma_0 C \bar D_{\mu L}^T (\bar x),
\end{array}
\right.
& &
\left\{ \begin{array}{rcl}
e_R (x) &\to& \gamma_0 C \bar e_R^T (\bar x),
\\*[1mm]
\mu_R (x) &\to& \gamma_0 C \bar \tau_R^T (\bar x),
\\*[1mm]
\tau_R (x) &\to& \gamma_0 C \bar \mu_R^T (\bar x),
\end{array}
\right.
\\*[-2mm]
CP: & &
\\*[-2mm]
& &
\left\{ \begin{array}{rcl}
\nu_{eR} (x) &\to& \gamma_0 C \bar \nu_{eR}^T (\bar x),
\\*[1mm]
\nu_{\mu R} (x) &\to& \gamma_0 C \bar \nu_{\tau R}^T (\bar x),
\\*[1mm]
\nu_{\tau R} (x) &\to& \gamma_0 C \bar \nu_{\mu R}^T (\bar x),
\end{array}
\right.
& &
\left\{
\begin{array}{rcl}
\sigma_e (x) &\to& \sigma_e (\bar x),
\\*[1mm]
\sigma_\mu (x) &\to& \sigma_\tau (\bar x),
\\*[1mm]
\sigma_\tau (x) &\to& \sigma_\mu (\bar x),
\\*[1mm]
\phi_k (x) &\to& \phi_k^\ast (\bar x) \ \, \forall k,
\end{array}
\right.
\end{array}
\ee
where $x = \left( t, \vec r \right)$ and $\bar x = \left( t, - \vec r \right)$.
This $CP$ symmetry renders
$y_1, y_2, \ldots, y_7,$ and $m$ real.
The mass $m'$ will then be real,
because the scalars $\sigma_\alpha$
are Hermitian fields,
hence their VEV $s$ is real.
Even if the $CP$ symmetry is spontaneously broken
by (relatively) complex $v_1$, $v_2$, and $v_3$,
the ensuing phases of $a$,
$b$,
and $c$ may be withdrawn from $M$ through appropriate rephasings
of the light-neutrino fields.
Thus,
\emph{there exists a restriction of our model in which $M$ is real}.

We proceed to fit the predictions of our model,
\textit{viz.}\ eqs.~(\ref{pred}),
to the phenomenological data.
The neutrino masses are $m_{1,2,3}$.
We use the standard parameterization of lepton mixing in ref.~\cite{pdg},
through three mixing angles $\theta_{12}$,
$\theta_{13}$,
and $\theta_{23}$ and one $CP$-violating phase $\delta$.
We have used the following allowed ranges for the various observables:
\be
\begin{array}{rcccl}
6.99 \times 10^{-5}\, \mbox{eV}^2 &\le& m_2^2 - m_1^2 &\le&
8.20 \times 10^{-5}\, \mbox{eV}^2, \\*[1mm]
2.16 \times 10^{-3}\, \mbox{eV}^2 &\le& \left| m_3^2 - m_1^2 \right| &\le&
2.74 \times 10^{-3}\, \mbox{eV}^2, \\*[1mm]
0.25 &\le& \sin^2{\theta_{12}} &\le& 0.37, \\*[1mm]
0.016 &\le& \sin^2{\theta_{13}} &\le& 0.033, \\*[1mm]
0.33 &\le& \sin^2{\theta_{23}} &\le& 0.68.
\end{array}
\ee
These ranges simultaneously encompass all the corresponding $3\sigma$ ranges
furnished by
the relevant phenomenological analyses~\cite{fogli,tortola,schwetz}.
We stress that,
even though the fit presented here uses these quite
ample ranges,
we have also found that
\emph{most observables easily fall within their respective
$1\sigma$ ranges}\/ given in,
for instance,
ref.~\cite{fogli};
the exception is the mixing angle $\theta_{12}$,
which is in our model rather large.
We have left $\delta$ free,
even though refs.~\cite{fogli,schwetz}
provide
some bounds on it,
which are,
however,
valid only at the $1 \sigma$ level.

\begin{figure}
\centerline{\epsfysize=7cm \epsfbox{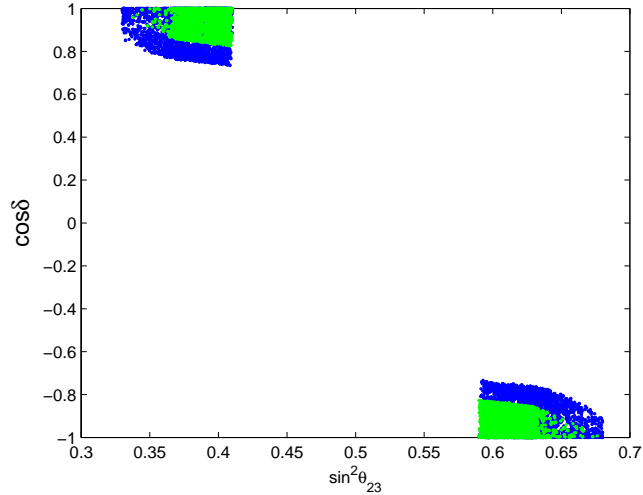}}
\caption{Scatter plot of $\cos{\delta}$ against $\sin^2{\theta_{23}}$.
The green (light grey) points are for an inverted neutrino mass spectrum
($m_3 < m_{1,2}$),
the blue (black) points for a normal one
($m_3 > m_{1,2}$); this convention is used in all the figures of this paper.}
\label{fig:mod1}
\end{figure}

\begin{figure}
\centerline{\epsfysize=7cm \epsfbox{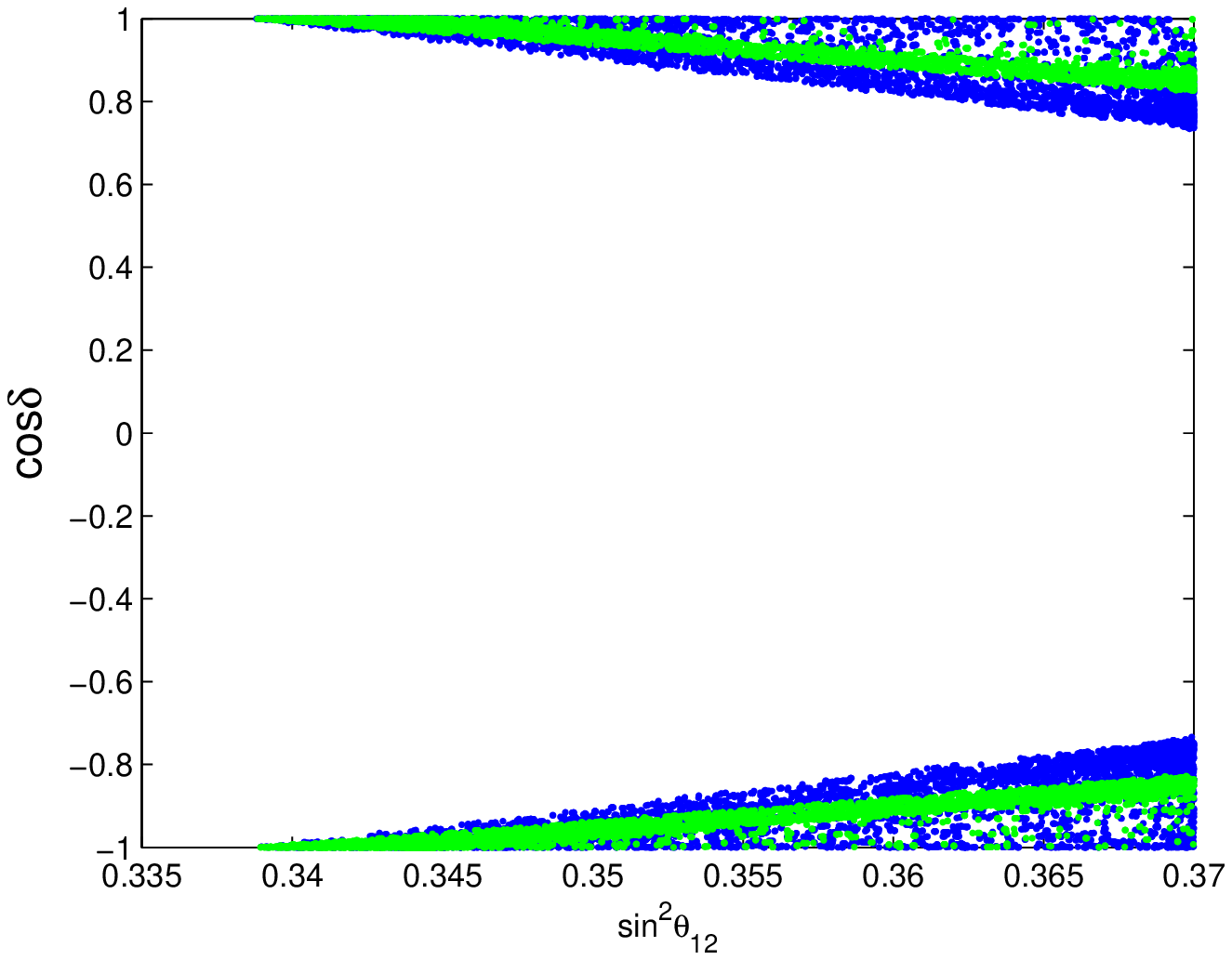}}
\caption{Scatter plot of $\cos{\delta}$ against $\sin^2{\theta_{12}}$.}
\label{fig:mod4}
\end{figure}

\begin{figure}
\centerline{\epsfysize=6cm \epsfbox{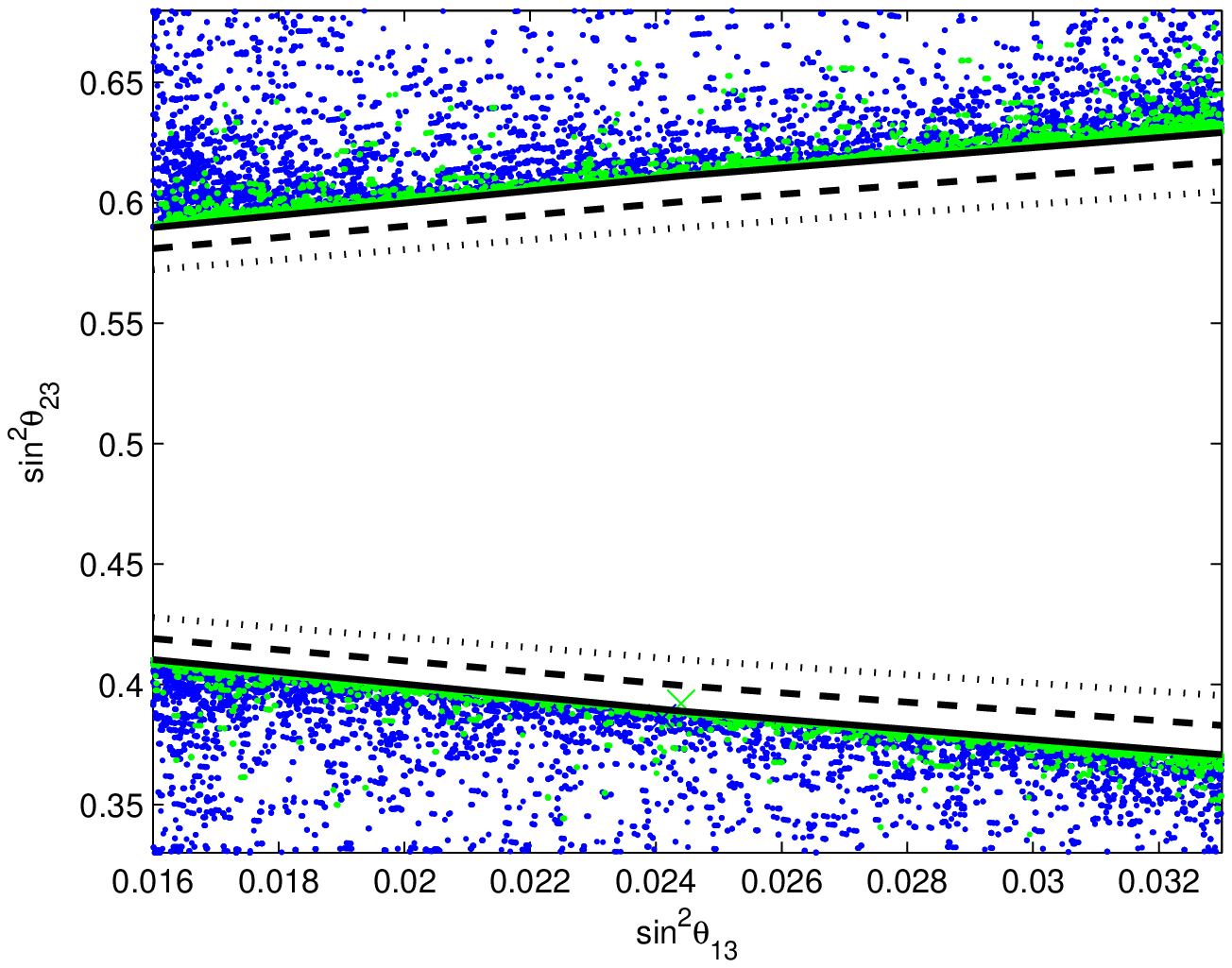}
\epsfysize=6cm \epsfbox{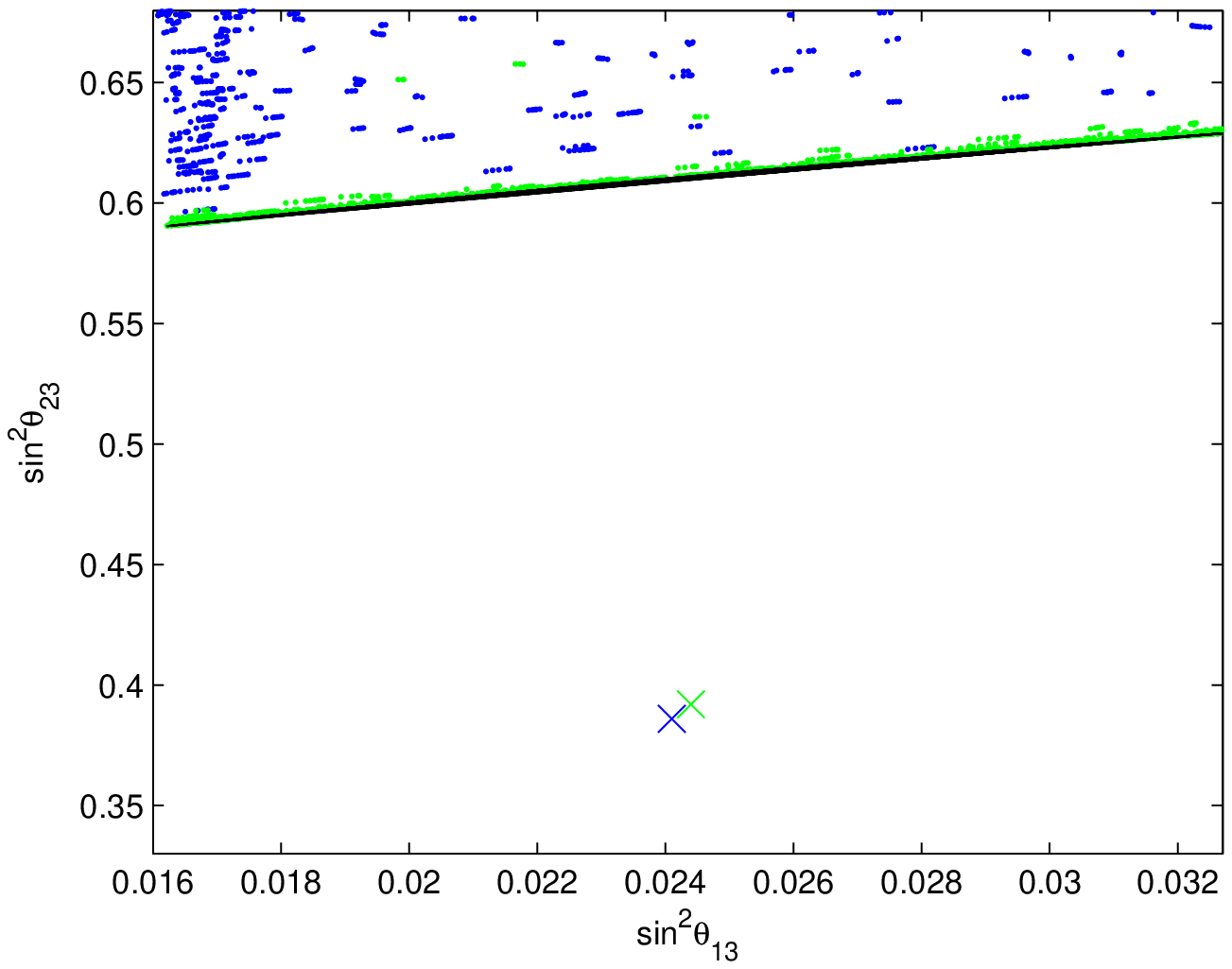}}
\caption{Scatter plots of $\sin^2{\theta_{23}}$ against $\sin^2{\theta_{13}}$.
The left plot is for the general version
and the right plot is for the real one with $\cos{\delta} = -1$.
The black lines correspond to
eq.~(\ref{costm2}),
predicted by TM$_2$~\cite{walter},
when $\cos{\delta} = \pm 1$ (full lines),
$\cos{\delta} = \pm 0.9$ (dashed lines),
and $\cos{\delta} = \pm 0.8$ (dotted lines).
The crosses
indicate the phenomenological best-fit points of ref.~\cite{fogli}.}
\label{fig:mod3}
\end{figure}

Our first finding is that in our model the phase $\delta$ must be close
to either $0$ or $\pi$;\footnote{The predictions of our model
are symmetric under $\mu \leftrightarrow \tau$.
In the parameterization of the lepton mixing matrix that we use,
the $\mu \leftrightarrow \tau$ interchange
corresponds to $\cos{\delta} \to - \cos{\delta}$ and
$\sin^2{\theta_{23}} \to 0.5 - \sin^2{\theta_{23}}$.
This symmetry is easily observable in figs.~\ref{fig:mod1} and \ref{fig:mod4}
and in the left panel of fig.~\ref{fig:mod3}.}
if $\delta \approx 0$ then $\theta_{23}$ is in the first octant
while $\theta_{23}$ is in the second octant
when $\delta \approx \pi$.\footnote{This contradicts
the phenomenological findings
(at the $1 \sigma$ level)
of ref.~\cite{fogli},
according to which $\theta_{23}$ lies in the first octant
and $\delta$ is close to $\pi$.
However,
those findings are not in agreement
with the ones of refs.~\cite{tortola,schwetz}.}
This can be seen in the scatter plot of fig.~\ref{fig:mod1}.
Therefrom one gathers that
in our model $\left| \cos{\delta} \right| \gtrsim 0.7$
($0.8$ if the neutrino mass spectrum is inverted)
and $\left| \sin^2{\theta_{23}} - 0.5 \right| \gtrsim 0.08$.

\begin{figure}[htb]
\centerline{\epsfysize=6cm \epsfbox{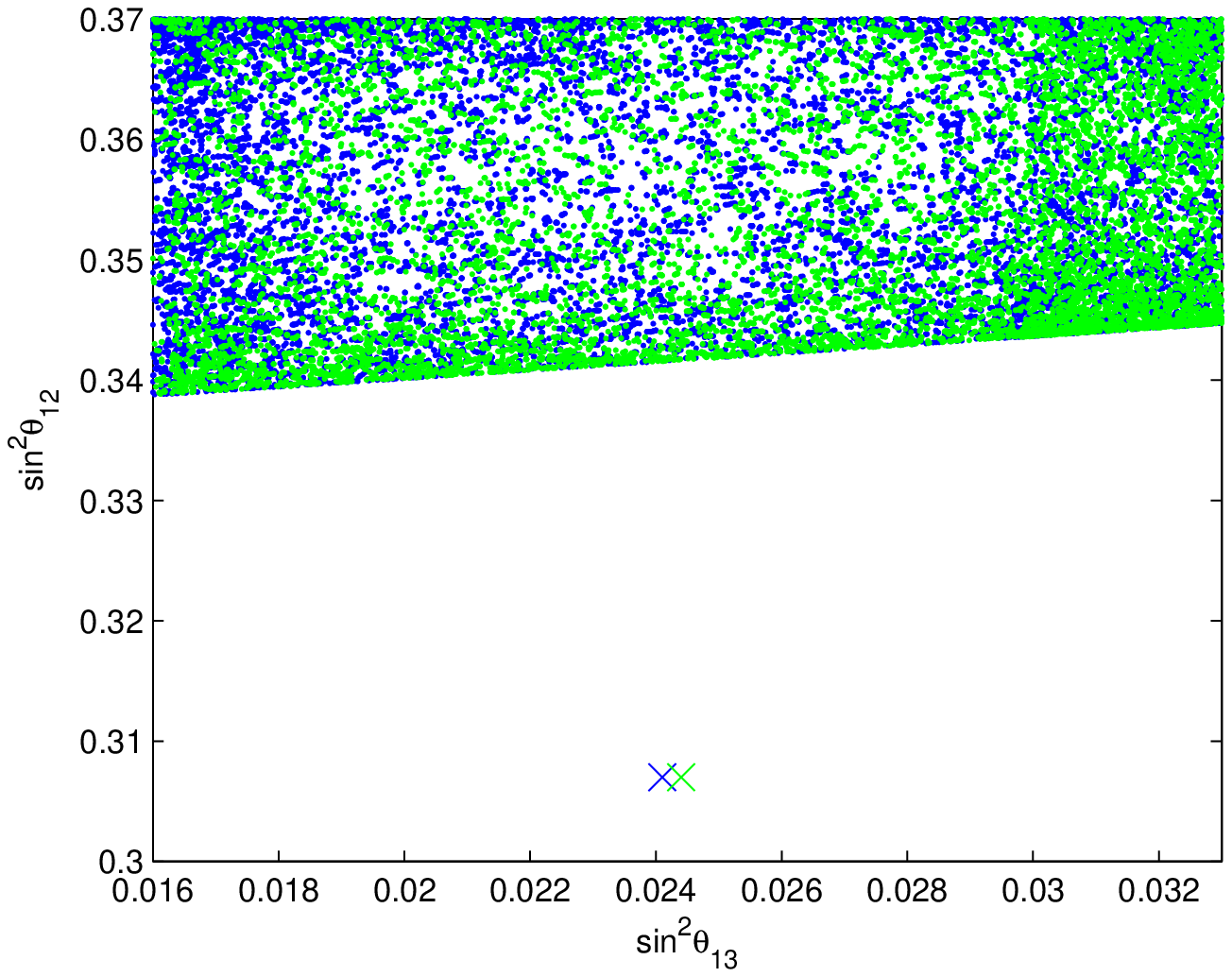}
\epsfysize=6cm \epsfbox{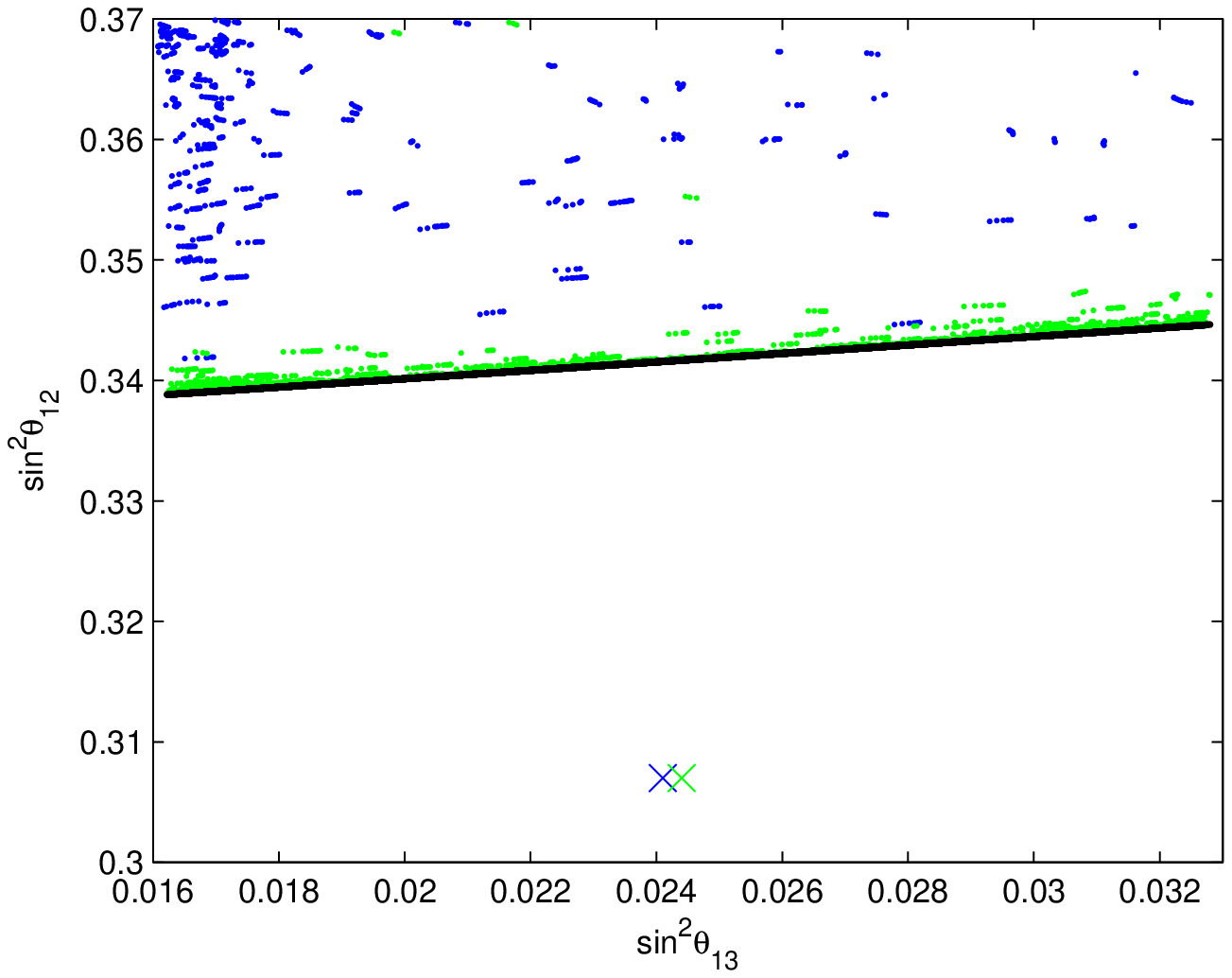}}
\caption{Scatter plots of $\sin^2{\theta_{12}}$ against $\sin^2{\theta_{13}}$.
The left plot is for the general (complex) version of the model,
the right plot is for the real version with $\cos{\delta} = -1$.
The black line in the right plot displays the prediction~\cite{walter}
$\sin^2{\theta_{12}} \left( 1 - \sin^{2}{\theta_{13}} \right) = 1/3$
of trimaximal mixing (TM$_2$).
The crosses
(blue (black) for a normal neutrino mass spectrum,
green (light grey) for an inverted spectrum)
indicate the phenomenological best-fit points of ref.~\cite{fogli}.}
\label{fig:mod2}
\end{figure}

In fig.~\ref{fig:mod2}
one sees that our model is unable to predict $\theta_{13}$
but neatly predicts $\sin^2{\theta_{12}} \gtrsim 0.34$.
This is a rather large value,
which is allowed by the phenomenological data only at the $2 \sigma$
(or even $3 \sigma$) level.
It can moreover be seen in fig.~\ref{fig:mod2}
(and also in fig.~\ref{fig:mod3})
that the restricted real version of our model
does not have much more predictive power than the
general version, even though it has one degree of freedom less.

In the right panel of fig.~\ref{fig:mod2} one observes that
the \emph{lower bound}\/ on $\theta_{12}$ in our model coincides
with the \emph{prediction}\/ for $\theta_{12}$
in a model with trimaximal mixing
(TM$_2$ in the nomenclature of ref.~\cite{werner}).
Trimaximal mixing is defined to be the situation where
$\left| U_{\alpha 2} \right| = 3^{-1/2}\, \ \forall
\alpha \in \left\{ e, \mu, \tau \right\}$.
A model with TM$_2$ has been suggested a few years ago~\cite{walter}.
In TM$_2$ $\left| U_{e2} \right|^2
= \sin^2{\theta_{12}} \cos^2{\theta_{13}} = 1/3$
and therefore $\sin^2{\theta_{12}} \approx 0.34$,
which is a bit large but has not deterred several authors---see for instance
ref.~\cite{memenga}---from having recently suggested
models and \textit{Ans\"atze}\/ featuring TM$_2$.
Note that TM$_2$,
just as our model,
is characterized by predictions for lepton mixing
which are invariant under any permutation of the lepton flavours.

One further prediction of TM$_2$ is\footnote{Reference~\cite{ge}
contains relations that generalize eq.~(\ref{costm2}).}
\be
\label{costm2}
\left( \cos{\delta} \right) \tan{\left( 2 \theta_{23} \right)}
= \frac{\cos{\left( 2 \theta_{13} \right)}}
{\left( \sin{\theta_{13}} \right)
\sqrt{3 \cos^2{\theta_{13}} - 1}} > 0.
\ee
Therefore,
in TM$_2$,
just as in our model,
$\delta$ is in the first (or fourth) quadrant
when $\theta_{23}$ is in the first octant,
and $\delta$ is in the second (or third) quadrant
when $\theta_{23}$ is in the second octant.
Moreover,
in TM$_2$ $\theta_{23}$ becomes closer to $\pi / 4$
when $\left| \cos{\delta} \right|$ becomes smaller.

Figure~\ref{fig:mod4}
shows that,
if we want a lower $\theta_{12}$ in our model,
then we must accept $\left| \cos{\delta} \right|$ to be ever closer to 1,
\textit{i.e.}\
a more stringent phenomenological
upper bound on $\sin^2{\theta_{12}}$
translates in our model into a more stringent lower bound
on $\left| \cos{\delta} \right|$.

In fig.~\ref{fig:mod3} one sees that our model's bound on $\theta_{23}$
depends only faintly on $\theta_{13}$.
One also sees that our model's \emph{lower bound}\/
on $\left| 0.5 - \sin^2{\theta_{23}} \right|$
coincides with the \emph{prediction}\/ of TM$_2$
for $\left| 0.5 - \sin^2{\theta_{23}} \right|$
when $\left| \cos{\delta} \right|$ is
\emph{maximal}.

\begin{figure}
\centerline{\epsfysize=6cm \epsfbox{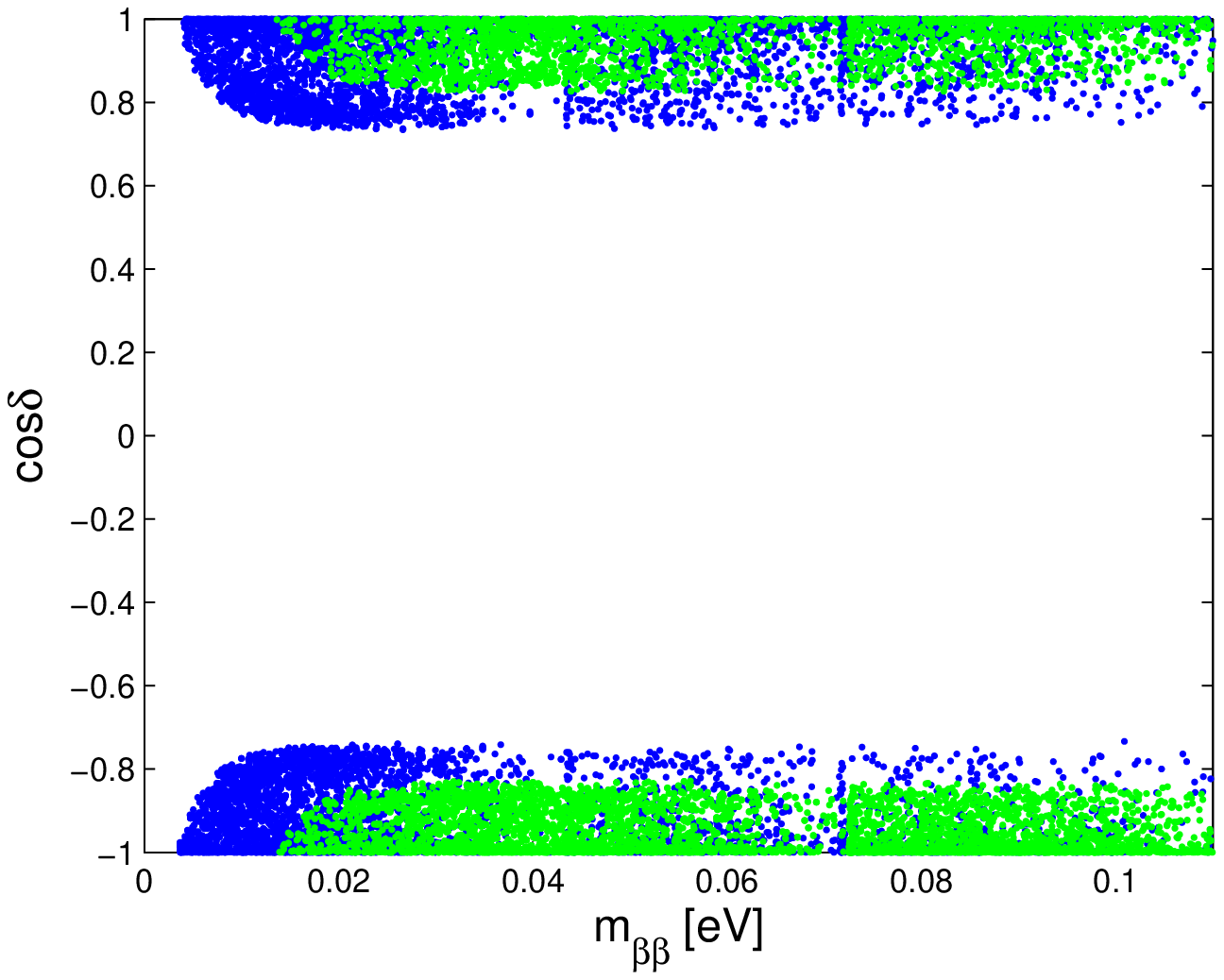}
\epsfysize=6cm \epsfbox{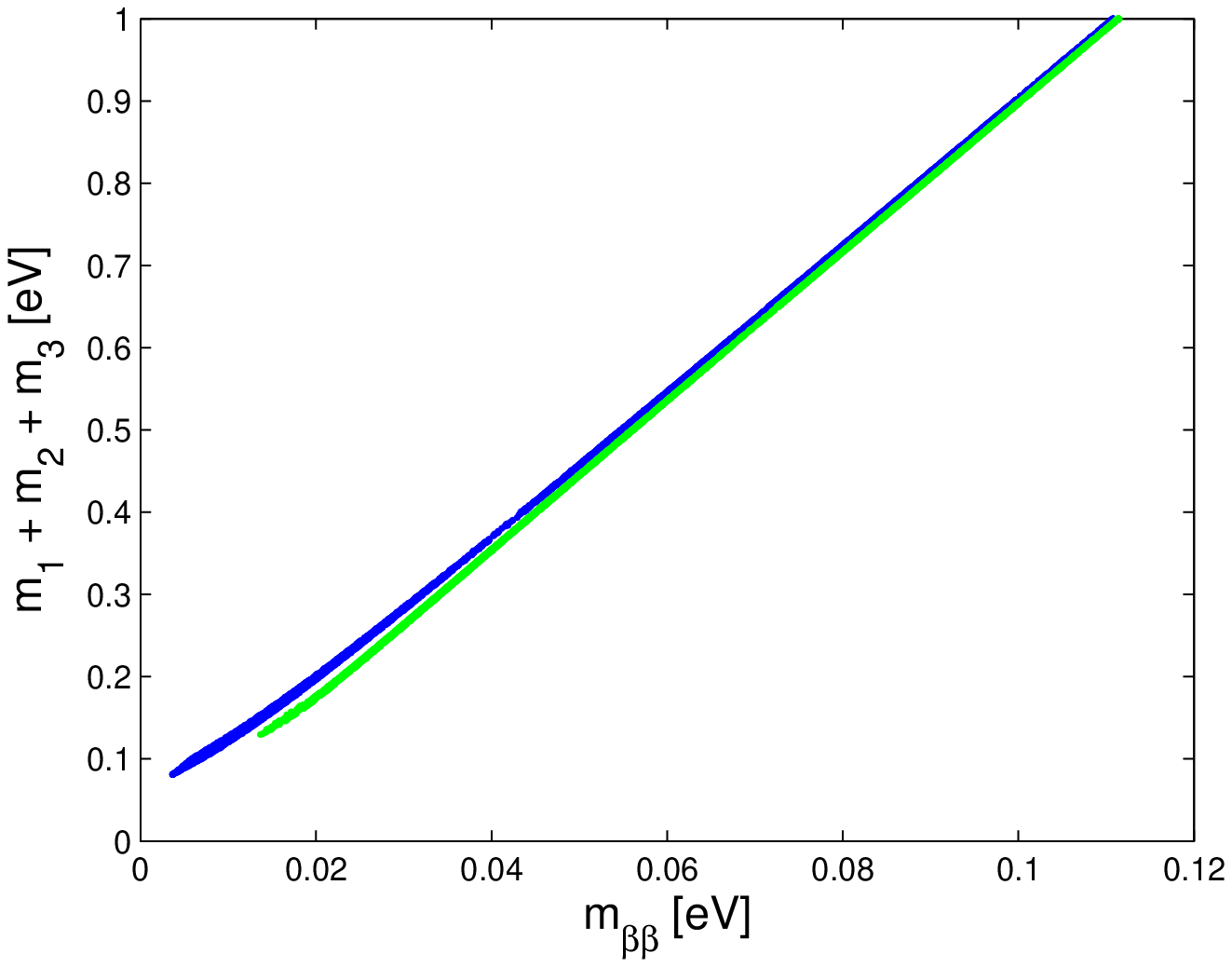}}
\caption{Ranges of allowed masses of the neutrinos in our model.
The left scatter plot shows
the neutrino mass relevant for neutrinoless double beta decay
as a function of $\cos{\delta}$;
the right plot correlates that quantity
with the sum of the three light neutrino masses.
We have imposed the cut $m_1 + m_2 + m_3 \le 1\, \text{eV}$
on the sum of the neutrino masses.}
\label{fig:mod7}
\end{figure}

Figure~\ref{fig:mod7}
shows the scale of the neutrino masses in our model.

We next speculate on possible experimental indications for
(or against)
our model.
As is clear in the first three lines of eq.~(\ref{yukawa}),
in our model the Yukawa couplings of the neutral scalars
to the charged leptons conserve flavour.
Therefore,
flavour-changing neutral Yukawa interactions
only arise at loop level and are suppressed by a loop factor
$\left( 16 \pi^2 \right)^{-1}$ and by two additional Yukawa couplings.
One may,
moreover,
show~\cite{grimus} that the
(loop induced)
flavour changing neutral couplings of the charged leptons
to the photon and to the $Z^0$ are suppressed by inverse powers
of the seesaw scale and are therefore,
in general,
unobservably small.
Therefore,
no decays like $\mu \to e \gamma$ or $Z^0 \to e^+ \mu^-$ are expected,
but decays like $h \to \tau^+ \mu^-$ might be observable at the LHC
($h$ is the observed scalar particle with mass 125\,GeV).

When extending our model to the quark sector,
one may either add to it further Higgs doublets or---a more
economic possibility---make the quarks have Yukawa couplings
to any one of the Higgs doublets $\phi_1$,
$\phi_2$,
or $\phi_3$
(or else the up-type quarks may couple to one of them
and the down-type quarks to another one).
The fact that our Higgs doublets are singlets of the flavour group
allows for this inviting possibility.
Depending on the specific Yukawa couplings used,
the signatures of the model at the LHC will vary.
It is worth pointing out that,
in any multi-Higgs-doublet model (MHDM),
a physical scalar couples to $Z^0 Z^0$
with a coupling at most as strong as
the one of the Higgs particle of the Standard Model
(SM)~\cite{ogreid}.
This naively suggests that the observed decay $h \to Z^0 Z^0$
could be used
to strongly constrain the parameter space of a MHDM,
but this is not the case.
Indeed,
in a MHDM the couplings of any particular scalar to the top and bottom quarks
might be either (much) stronger or (much) weaker than in the SM,
and therefore in a MHDM
both the production cross section and the total decay width
of $h$ will in general be at variance with those expected in the SM.

To summarize,
we have presented in this paper a seesaw model
featuring a simple application of the $A_4$ symmetry
and which makes the predictions~(\ref{pred}) for lepton mixing.
We have shown that our model is compatible with experiment
provided $\theta_{12}$ is in the upper part of its allowed range.
Our model predicts non-maximal $\theta_{23}$
and also makes the prediction that $\left| \cos{\delta} \right|$ is very close to 1.

\vspace*{5mm}

\section*{Acknowledgements:}
LL thanks Jo\~ao P.~Silva for a useful discussion.
The work of PMF is supported by
the Portuguese Foundation for Science and Technology (FCT)
through the projects PEst-OE/FIS/UI0618/2011,
PTDC/FIS/117951/2010,
and the FP7 Reintegration Grant PERG08-GA-2010-277025.
The work of LL is supported through
the Marie Curie Initial Training Network ``UNILHC'' PITN-GA-2009-237920
and through the projects PEst-OE/FIS/UI0777/2013,
PTDC/FIS-NUC/0548/2012,
and CERN/FP/123580/2011 of FCT.
The work of POL is supported by
the Austrian Science Fund (FWF) through the project P~24161-N16.

\vspace*{3mm}

\begin{appendix}

\setcounter{equation}{0}
\renewcommand{\theequation}{A\arabic{equation}}

\section{Appendix: the VEVs of the $\sigma_\alpha$}

The potential for the $\sigma_\alpha$ is
\ba
V_\sigma &=&
\mu \left( \sigma_e^2 + \sigma_\mu^2 + \sigma_\tau^2 \right)
+ \lambda_1 \left( \sigma_e^2 + \sigma_\mu^2 + \sigma_\tau^2 \right)^2
\nonumber \\ & &
+ \tilde m\, \sigma_e \sigma_\mu \sigma_\tau
+ \lambda_2 \left( \sigma_e^2 \sigma_\mu^2 + \sigma_\mu^2 \sigma_\tau^2
+ \sigma_\tau^2 \sigma_e^2 \right).
\ea
We have neglected terms which include both the $\sigma_\alpha$ and the $\phi_k$
since the VEVs of the $\phi_k$
should be much smaller than the VEVs of the $\sigma_\alpha$
and therefore those terms should have a negligible influence
on the equations which determine the VEVs of the $\sigma_\alpha$.

One may parameterize the VEVs of the $\sigma_\alpha$ as
\be
\left\langle 0 \left| \sigma_e \right| 0 \right\rangle = U \cos{\vartheta},
\quad
\left\langle 0 \left| \sigma_\mu \right| 0 \right\rangle
= U \sin{\vartheta} \cos{\varphi},
\quad
\left\langle 0 \left| \sigma_\tau \right| 0 \right\rangle
= U \sin{\vartheta} \sin{\varphi},
\ee
with $U \ge 0$,
$\vartheta \in \left[ 0, \pi \right]$,
and $\varphi \in \left[ 0, 2 \pi \right[$.
Then,
\ba
V_{\sigma 0} \equiv \left\langle 0 \left| V_\sigma \right| 0 \right\rangle &=&
\mu U^2 + \lambda_1 U^4
+ \frac{\tilde m U^3}{2}\, \sin^2{\vartheta} \cos{\vartheta} \sin{2 \varphi}
\nonumber \\ & &
+ \lambda_2 U^4 \sin^2{\vartheta} \left(
\cos^2{\vartheta} + \frac{1}{4}\, \sin^2{\vartheta} \sin^2{2 \varphi} \right).
\ea
There is a range of $\tilde m$ and $\lambda_2$
for which the minimum of $V_{\sigma 0}$ occurs when
$\sin{2 \varphi}$ is at the boundary of its range,
\textit{viz.}\ when $\sin{2 \varphi} = 1$.
There,
\be
V_{\sigma 0} =
\mu U^2 + \lambda_1 U^4
+ \frac{\tilde m U^3}{2}\, \sin^2{\vartheta} \cos{\vartheta}
+ \frac{\lambda_2 U^4}{4}\, \sin^2{\vartheta} \left(
1 + 3 \cos^2{\vartheta} \right).
\ee
Then,
\be
\frac{\partial V_{\sigma 0}}{\partial \cos{\vartheta}} =
U^3 \left( 1 - 3 \cos^2{\vartheta} \right)
\left( \frac{\tilde m}{2} + \lambda_2 U \cos{\vartheta} \right).
\ee
Within a range of $\tilde m$ and $\lambda_2$ the minimum of $V_{\sigma 0}$
occurs when $\cos^2{\vartheta} = 1/3$ and,
indeed,
$\cos{\vartheta} = 1 \left/ \sqrt{3} \right.$.
In this way one obtains
$\left\langle 0 \left| \sigma_e \right| 0 \right\rangle
= \left\langle 0 \left| \sigma_\mu \right| 0 \right\rangle
= \left\langle 0 \left| \sigma_\tau \right| 0 \right\rangle
= U \! \left/ \sqrt{3} \right.$ as desired.

\end{appendix}


\begin{thebibliography}{99}

\bibitem{nonzero}
Y.~Abe \textit{et al.} (Double Chooz Coll.),
\textit{Indication for the disappearance of reactor electron antineutrinos
in the Double Chooz experiment},
\textit{Phys.\ Rev.\ Lett.}\ \textbf{108} (2012) 131801;
\\
F.P.~An \textit{et al.} (Daya Bay Coll.),
\textit{Observation of electron-antineutrino disappearance at Daya Bay},
\textit{Phys.\ Rev.\ Lett.}\ \textbf{108} (2012) 171803;
\\
J.K.\ Ahn \textit{et al.} (RENO Coll.),
\textit{Observation of reactor electron antineutrino disappearance in
the RENO experiment},
\textit{Phys.\ Rev.\ Lett.}\ \textbf{108} (2012) 191802.

\bibitem{flavons}
F.~Feruglio, C.~Hagedorn, Y.~Lin, and L.~Merlo,
\textit{Lepton flavour violation in a supersymmetric model with $A_4$
flavour symmetry},
\textit{Nucl.\ Phys.\ B}\/ \textbf{832} (2010) 251;
\\
S.F.~King and C.~Luhn,
\textit{$A_4$ models of tri-bimaximal-reactor mixing},
\textit{J.\ High Energy Phys.}\ \textbf{1203} (2012) 036;
\\
M.-C.~Chen, J.~Huang, J.-M.~O'Bryan, A.M.~Wijangco, and F.~Yu,
\textit{Compatibility of $\theta_{13}$ and the type I seesaw model
with $A_4$ symmetry},
\textit{J.\ High Energy Phys.}\ \textbf{1302} (2013) 021;
\\
M.~Holthausen, M.~Lindner, and M.A.~Schmidt,
\textit{Lepton flavor at the electroweak scale: A complete $A_4$ model},
\textit{Phys.\ Rev.\ D}\/ \textbf{87} (2013) 033006;
\\
S.F.~King, S.~Morisi, E.~Peinado, and J.W.F.~Valle,
\textit{Quark--lepton mass relation in a realistic $A_4$ extension
of the Standard Model},
\textit{Phys.\ Lett.\ B} {\bf 724} (2013) 68;
\\
S.~Morisi, M.~Nebot, K.M.~Patel, E.~Peinado, and J.W.F.~Valle,
\textit{Quark--lepton mass relation and CKM mixing
in an $A_4$ extension of the Minimal Supersymmetric Standard Model},
\textit{Phys.\ Rev.\ D} {\bf 88}, 036001 (2013).

\bibitem{renormalizable}
D.~Meloni, S.~Morisi, and E.~Peinado,
\textit{Neutrino phenomenology and stable dark matter with $A_4$},
\textit{Phys.\ Lett.\ B}\/ \textbf{697} (2011) 339;
\\
S.~Gupta, A.S.~Joshipura and K.M.~Patel,
\textit{Minimal extension of tri-bimaximal mixing and generalized $Z_2 \times Z_2$ symmetries},
\textit{Phys.\ Rev.\ D}\/ \textbf{85} (2012) 031903;
\\
H.~Ishimori and E.~Ma,
\textit{New simple $A_4$ neutrino model
for nonzero $\theta_{13}$ and large $\delta_{CP}$},
\textit{Phys.\ Rev.\ D}\/ \textbf{86} (2012) 045030;
\\
E.~Ma,
\textit{Self-organizing neutrino mixing matrix},
\textit{Phys.\ Rev.\ D}\/ \textbf{86} (2012) 117301;
\\
E.~Ma, A.~Natale, and A.~Rashed,
\textit{Scotogenic $A_4$ neutrino model for nonzero $\theta_{13}$
and large $\delta_{CP}$},
\textit{Int.\ J.\ Mod.\ Phys.\ A}\/ \textbf{27} (2012) 1250134;
\\
Y.~Ben~Tov, X.-G.~He, and A.~Zee,
\textit{An $A_4 \times \mathbb{Z}_4$ model for neutrino mixing},
\textit{J.\ High Energy Phys.}\ {\bf 1212} (2012) 093;
\\
Y.H.~Ahn, S.K.~Kang, and C.S.~Kim,
\textit{Spontaneous $CP$ violation in $A_4$ flavor symmetry and leptogenesis},
\textit{Phys.\ Rev.\ D} {\bf 87} (2013) 113012;
\\
A.E.~Carcamo Hernandez, I.d.M.~Varzielas, S.G.~Kovalenko, H.~P\"as and I.~Schmidt,
\textit{Lepton masses and mixings in a $A_{4}$ multi-Higgs model with radiative seesaw},
arXiv:1307.6499 [hep-ph].

\bibitem{mr}
K.S.~Babu, E.~Ma, and J.W.F.~Valle,
\textit{Underlying $A_4$ symmetry for the neutrino mass matrix
and the quark mixing matrix},
\textit{Phys.\ Lett.\ B}\/ \textbf{552} (2003) 207;
\\
For a modern version, see
S.~Morisi, D.V.~Forero, J.C.~Rom\~ao, and J.W.F.~Valle,
\textit{Neutrino mixing with revamped $A_4$ flavour symmetry},
\textit{Phys.\ Rev.\ D} {\bf 88} (2013) 016003.

\bibitem{we}
P.M.~Ferreira, L.~Lavoura, and P.O.~Ludl,
\textit{Five models for lepton mixing},
\textit{J.\ High Energy Phys.}\ {\bf 1308} (2013) 113.

\bibitem{FN}
C.D.~Froggatt and H.B.~Nielsen,
\textit{Hierarchy of quark masses, Cabibbo angles and $CP$ violation},
\textit{Nucl.\ Phys.\ B}\/ {\bf 147} (1979) 277.

\bibitem{radovcic}
W.~Grimus, L.~Lavoura, and B.~Radov\v{c}i\'c,
\textit{Type~II seesaw mechanism for Higgs doublets
and the scale of new physics},
\textit{Phys.\ Lett.\ B}\/ \textbf{674} (2009) 117.

\bibitem{pdg}
J.~Beringer \textit{et al.}\ (Particle Data Group),
\textit{Review of particle physics},
\textit{Phys.\ Rev.\ D}\/ \textbf{86} (2012) 010001.

\bibitem{fogli}
G.L.~Fogli, E.~Lisi, A.~Marrone, D.~Montanino, A.~Palazzo, and A.M.~Rotunno,
\textit{Global analysis of neutrino masses, mixings and phases:
Entering the era of leptonic $CP$ violation searches},
\textit{Phys.\ Rev.\ D}\/ \textbf{86} (2012) 013012.

\bibitem{tortola}
D.V.~Forero, M.~T\'ortola, and J.W.F.~Valle,
\textit{Global status of neutrino oscillation parameters
after Neutrino--2012},
\textit{Phys.\ Rev.\ D}\/ \textbf{86} (2012) 073012.

\bibitem{schwetz}
M.C.~Gonzalez-Garcia, M.~Maltoni, J.~Salvado, and T.~Schwetz,
\textit{Global fit to three neutrino mixing:
Critical look at present precision},
\textit{J.\ High Energy Phys.}\ \textbf{1212} (2012) 123.

\bibitem{werner}
C.H.~Albright, A.~Dueck, and W.~Rodejohann,
\textit{Possible alternatives to tri-bimaximal mixing},
\textit{Eur.\ Phys.\ J.\ C}\/ {\bf 70} (2010) 1099.

\bibitem{walter}
W.~Grimus and L.~Lavoura,
\textit{A model for trimaximal lepton mixing},
\textit{J.\ High Energy Phys.}\ {\bf 0809} (2008) 106.

\bibitem{memenga}
N.~Memenga, W.~Rodejohann, and H.~Zhang,
\textit{$A_4$ flavor symmetry model for Dirac neutrinos and sizable $U_{e3}$},
\textit{Phys.\ Rev.\ D}\/ {\bf 87} (2013) 053021;
\\
S.F.~King, T.~Neder, and A.J.~Stuart,
\textit{Lepton mixing predictions from $\Delta ( 6 n^2 )$
family symmetry},
arXiv:1305.3200 [hep-ph];
\\
H.~Qu and B.-Q.~Ma,
\textit{New mixing pattern for neutrinos},
\textit{Phys.\ Rev.\ D} {\bf 88} (2013) 037301.

\bibitem{ge}
S.-F.~Ge, D.A.~Dicus, and W.W.~Repko,
\textit{$\mathbb{Z}_2$ symmetry prediction for the leptonic Dirac $CP$ phase},
\textit{Phys.\ Lett.\ B}\/ \textbf{702} (2011) 220;
\\
S.-F.~Ge, D.A.~Dicus, and W.W.~Repko,
\textit{Residual symmetries for neutrino mixing with a large $\theta_{13}$
and nearly maximal $\delta_D$},
\textit{Phys.\ Rev.\ Lett.}\ \textbf{108} (2012) 041801.

\bibitem{grimus}
W.~Grimus and L.~Lavoura,
\textit{Soft lepton flavor violation in a multi-Higgs-doublet seesaw model},
\textit{Phys.\ Rev.\ D}\/ \textbf{66} (2002) 014016;
\\
R.~Alonso, M.~Dhen, M.B.~Gavela, and T.~Hambye,
\textit{Muon conversion to electron in nuclei in type-I seesaw models},
\textit{J.\ High Energy Phys.}\ \textbf{1301} (2013) 118.

\bibitem{ogreid}
See for instance
W.~Grimus, L.~Lavoura, O.M.~Ogreid, and P.~Osland,
\textit{A precision constraint on multi-Higgs-doublet models},
\textit{J.\ Phys.\ G}\/ \textbf{35} (2008) 075001;
\\
G.C.~Branco, P.M.~Ferreira, L.~Lavoura, M.N.~Rebelo, M.~Sher, and J.P.~Silva,
\textit{Theory and phenomenology of two-Higgs-doublet models},
\textit{Phys.\ Rep.}\ \textbf{516} (2012) 1.

\end{thebibliography}
\end{document}